\documentclass[12pt]{article}
\usepackage{amsmath,amsfonts,amssymb}
\newcommand{\nn}{\noindent}
\newcommand{\no}{\nonumber\\}
\newcommand{\be}{\begin{equation}}
\newcommand{\ee}{\end{equation}}
\newcommand{\ba}{\begin{eqnarray}}
\newcommand{\ea}{\end{eqnarray}}

\newcommand{\la}[1]{\label{#1}}

\newcommand{\gl}[1]{(\ref{#1})}

\date{}
\begin{document}
\title{ Non-linear Supersymmetry for non-Hermitian,
non-diagonalizable Hamiltonians: I. General properties}
\author{
A. A. Andrianov$^{\dag\star}$,\,F. Cannata$^{\star}$,\,A. V. Sokolov $^{\dag}$\\
{$^\dag$ \it  V.A.Fock Institute of
Physics, Sankt-Petersburg State University,}\\
{$^\star$ \it Dipartimento di Fisica and INFN Bologna, Italy}} \maketitle

\abstract{
 We study complex potentials and related non-diagonalizable
 Hamiltonians
with special emphasis on formal definitions of
  associated functions and Jordan cells.
 The non-linear SUSY for complex potentials is considered and
the  theorems characterizing its structure are presented.  We define the class of complex potentials 
invariant under SUSY transformations for (non-)diagonalizable Hamiltonians and formulate several results 
concerning the properties of associated functions . We comment on the applicability of these results for 
{\it softly} non-Hermitian PT-symmetric Hamiltonians. The role of SUSY (Darboux) transformations in 
increasing/decreasing of Jordan cells in SUSY partner Hamiltonians is thoroughly analyzed and summarized in 
the Index Theorem.  The properties of non-diagonalizable Hamiltonians as well as the Index Theorem are 
illustrated in the solvable examples of non-Hermitian reflectionless Hamiltonians . The rigorous proofs are 
relegated to the Part II of this paper. At last, some peculiarities in resolution of identity for discrete 
and continuous
  spectra with a zero-energy bound state at threshold are discussed.
}

\section{Introduction}

Quantum Physics of open systems often deals with incomplete information on the influence of an environment 
and can be adequately described by non-Hermitian Hamiltonians with a non-positive imaginary part. This kind 
of effective description has been employed in Condensed Matter, Quantum Optics and Hadronic and Nuclear 
Physics \cite{optical} -- \cite{muga} for many years. Non-self-adjoint operators were also under 
mathematical investigations \cite{dunf,pavl,davies} and recently interesting examples of non-Hermitian 
effective Hamiltonian operators have been found for the quantum many-body equations \cite{yak}.

The PT-symmetric Quantum Mechanics proposed in \cite{bender,bender1} and developed in 
\cite{bender1}--\cite{ventura} and its pseudo-Hermitian generalization \cite{mostaf,mostaf1,levai} 
describes a variety of non-Hermitian  Hamiltonians with real spectrum (but not all Hamiltonians with real 
spectrum are PT-symmetric \cite{complex0,acdicom}). There is a progress  in understanding some 
non-Hermitian but PT-symmetric Hamiltonians in terms of Krein spaces \cite{tanaka+}. This kind of Quantum 
Mechanics has attracted much interest as it may open the way to give a solid probabilistic interpretation 
of non-Hermitian dynamics by means of a positive pseudo-norm \cite{complex2,bender+}. PT-symmetry endows 
with a physical meaning the energy spectrum of some Hamiltonians formally unbounded from below 
\cite{bender,bender1}. The latter possibility for anharmonic oscillators with potentials unbounded from 
below was observed long ago \cite{andr,buslaev} but only recently has been associated with a PT-symmetry 
\cite{jones}.

For complex, non-Hermitian potentials the natural spectral decomposition involves biorthogonal states 
\cite{curt}. Moreover the Hamiltonians may not be diagonalizable \cite{solov} but can be reduced only to a 
quasi-diagonal form with a number of Jordan cells \cite{mostaf}.  This feature  appears at level crossing 
which, in fact, occurs under specific circumstances in atomic and molecular spectra \cite{solov} and Optics 
\cite{berry,mois2} as well as in PT-symmetric quantum systems \cite{dorey,znojil,sams1}. There are also 
certain links \cite{gadella} to the occurrence of non-Hermitian degeneracies for essentially Hermitian 
Hamiltonians where the description has been developed  for complex eigenvalue Gamow states (resonances) 
unbounded in their  asymptotics and, in general, not belonging to the Hilbert space of physical wave 
functions. On the contrary, in what follows we  examine non-Hermitian Hamiltonians with normalizable bound  
and associated states. The subtleties of biorthogonality (the phenomenon of 
"self-orthogonality"\cite{mois2}) in resolution of identity and in definition of quantum averages of 
observables have been thoroughly analyzed in our paper \cite{sokancan} .

We find it certainly interesting and important to investigate the possible ways for quantum design  of such 
non-Hermitian quantum systems and in particular to extend the methods of non-linear SUSY algebra 
\cite{ais}--\cite{ddt} in order to keep under control the emerging of non-diagonal parts of those systems. 
In making a link to PT-symmetric systems  we restrict ourselves with a {\it soft} type of non-Hermiticity 
when the real part of a potential dominates over the imaginary one at both infinities and asymptotically 
such a potential remains {\it bounded} from below. Respectively the energy spectrum of a related system 
contains a number of bound states and possibly a continuum part bounded from below. Thus having in mind the 
SUSY quantum design one can, for instance, think of a chain of complex Hamiltonians produced by 
Darboux-Crum transformations from a real one as a good representative of the class of softly non-Hermitian 
systems. The general relations and theorems presented in Sections 2 -- 4 certainly hold also for 
PT-symmetric potentials with fixed asymptotics of ratio of imaginary and real part
({\it semihard} non-Hermiticity), say, for potentials with leading asymptotics $\lambda x^{2n} 
(ix)^\epsilon;\ \lambda >0 $ at infinities provided that the boundary conditions for eigenvalue problem do 
not require to move to complex coordinates \cite{bender1, dorey}({\it i.e.} for $|\epsilon| < 1$) . However 
the Lemmas and Theorem of Sections 5 and 6 are proven for softly non-Hermitian potentials and we pay hopes 
to extend them also on semihard non-Hermitian potentials in a nearest future.

We start in Sec.2 with the definitions and a summary of properties of non-Hermitian diagonalizable 
Hamiltonians and introduce the relevant  biorthogonal expansions. Then we consider non-diagonalizable 
non-Hermitian Hamiltonians with discrete spectrum and finite-size Jordan cells and discuss the choice of 
the biorthogonal basis with diagonal resolution of identity. The novel result of this section is the proof 
that a biorthogonal basis always exists which is made of a set of eigenfunctions and associated functions 
of the initial Hamiltonian and a set of their complex conjugates for the Hermitian conjugated Hamiltonian. 
Moreover if the Hamiltonian is PT-symmetric and this symmetry is not spontaneously broken on states 
(eigenvalues are real) then the elements of direct and conjugated bases are related by PT-reflection. In 
Sec.3  the origin of non-diagonalizable Hamiltonians is clarified to be level confluence.

Non-linear SUSY in QM is summarized and extended to complex potentials in Sec.4 with an emphasis to the 
possibility of conservation of PT-symmetry. Herein the important theorem on the polynomial structure of 
SUSY algebra with transposition symmetry  as well as the strip-off theorem describing the minimization of 
the differential order  of intertwining operators are adapted to the complex potentials. These theorems 
involve the zero-mode subspaces of supercharge components -- intertwining operators and their mapping by 
Hamiltonians -- matrices $S$. They are well compatible with PT-symmetry (if any). The relationship between 
superpotentials and Wronskians involving associated functions is discussed.

In Sec.5 we present the class of complex potentials invariant under SUSY  transformations for 
(non-)diagonalizable Hamiltonians: this class covers the systems with soft breaking of Hermiticity and 
essentially real continuum spectrum (if any). For this case we formulate several results characterizing the 
normalizability of associated functions at $+\infty$ and/or $-\infty$. The necessary conditions for SUSY 
transformation functions are found to provide a pre-planned Jordan structure of a SUSY partner Hamiltonian. 
These results allow to unravel the relation between Jordan cells in SUSY partner Hamiltonians the latter 
being described by the Index Theorem in Sec 6. It represents the main result of the present paper. The 
Index Theorem relates the dimensions of Jordan cells of super-partner Hamiltonians at any energy level with 
characteristics of intertwining operator kernels (matrices $S$)and in fact exhaustively describes the 
quantum design options for softly non-Hermitian Hamiltonians.  Needless to say that the latter theorem is 
also compatible with PT-symmetry when non-Hermiticity , for instance, is introduced into P-even potentials 
by shifting of coordinates into complex plane (see examples in \cite{znojil}) . The illustration of 
properties of non-diagonalizable Hamiltonians as well as of the Index Theorem is thoroughly performed in 
Sec. 7 by the solvable example of non-Hermitian reflectionless Hamiltonians originated by SUSY 
trans\-for\-ma\-tions from the free particle Hamiltonian. The arising of non-diagonalizability is 
illuminated by an exactly solvable system with two coalescing bound states. In Conclusions we outline 
possible peculiarities of non-Hermitian Hamiltonians with continuous spectrum. The approaching to the 
continuum threshold yields more subtle  problems with normalizable eigen- and associated functions in 
continuum which may have zero binorm. As a consequence it may cause serious problems with the resolution of 
identity investigated in detail elsewhere \cite{sokancan}.

All the new results on the structure of non-diagonalizable  SUSY Hamiltonians presented in this (part of) 
paper are rigorously proved and justified in the accompanying (second part of) paper \cite{sokolov}.

\section{Non-Hermitian diagonalizable vs. non-dia\-go\-na\-lizable
Hamiltonians and biorthogonal expansions} In our paper we deal with complex  one-dimensional potentials 
$V(x) \not= V^*(x)$ and respectively with non-Hermitian Hamiltonians $h$ of Schr\"odinger 
type\footnote{Conventionally the system of units $m=1/2,\ \hbar = c =1$ will be used with dimensionless 
energies, momenta and coordinates.}, defined on the real axis, \be h \equiv - \partial^2 + V(x),\ee which 
are assumed to be symmetric or self-transposed under the $^t$ -- transposition operation, $h = h^t $. The 
notation $\partial \equiv d/dx$ is employed. Only scalar local potentials will be analyzed which are 
obviously symmetric under transposition (for some matrix non-diagonalizable problems, see 
\cite{mois1,matr}). Taking into account possible applications in PT-symmetric QM we specify complex 
potentials to give a {\it semihard} non-Hermiticity when the real part Re$V$ is bounded from below and the  
ratio Im$V/$Re$V$ remains finite and sufficiently small for large $x \rightarrow \pm\infty$ . In this case 
the eigenvalue problem can be safely posed keeping the boundary conditions at  $x \pm\infty$ on the real 
axis. Later on, in last two sections we restrict ourselves with {\it softly} non-Hermitian potentials with 
vanishing asymptotic ratios  Im$V/$Re$V =  o(1)$ .

Let us first define a class of one-dimensional non-Hermitian  {\it diagonalizable}
Hamiltonians $h$ with discrete spectrum  such  that:\\
a)  a biorthogonal system $\{|\psi_n\rangle, |\tilde\psi_n\rangle\}$ exists, \be
 h|\psi_n\rangle = \lambda_n|\psi_n\rangle,\qquad h^\dag|\tilde\psi_n\rangle=
\lambda_n^*|\tilde\psi_n\rangle,\qquad\langle\tilde\psi_n|\psi_m\rangle= 
\langle\psi_m|\tilde\psi_n\rangle=\delta_{nm},\ee b) the complete resolution of identity in terms of these  
bases and the spectral decomposition of the Hamiltonian
 hold (in the case of PT-symmetric potentials the necessary conditions for that are formulated in 
\cite{tanaka+}),
\be I=\sum\limits_n|\psi_n\rangle\langle\tilde\psi_n|,\qquad h= 
\sum\limits_n\lambda_n|\psi_n\rangle\langle\tilde\psi_n|.\ee

In the coordinate representation, \be \psi_n(x)=\langle x|\psi_n\rangle,\qquad\tilde\psi_n(x)=\langle x| 
\tilde\psi_n\rangle ,\ee the resolution of identity has the form, \be \delta(x-x')=\langle 
x'|x\rangle=\sum\limits_n\psi_n(x')\tilde\psi_n^*(x).\ee The differential equations, \be 
h\psi_n=\lambda_n\psi_n, \qquad h^\dag\tilde\psi_n=\lambda_n^*\tilde\psi_n , \ee and the fact that there is 
only one normalizable eigenfunction of $h$ for the eigenvalue $\lambda_n$ (up to a constant factor), allow 
one to conclude that \be \tilde\psi_n^*(x)\equiv\alpha_n\psi_n(x),\qquad\alpha_n={\rm{Const}}\ne0.\ee Hence 
the  system  $\{|\psi_n\rangle, |\tilde\psi_n\rangle\}$ can be redefined \be 
|\psi_n\rangle\to{1\over\sqrt{\alpha_n}}|\psi_n\rangle,\qquad 
|\tilde\psi_n\rangle\to\sqrt{\alpha_n^*}|\tilde\psi_n\rangle , \label{normal} \ee so that 
\be\tilde\psi_n^*(x)\equiv\psi_n(x),\qquad\int\limits_{-\infty}^{+\infty}\psi_n(x) 
\psi_m(x)\,dx=\delta_{nm}. \la{bior}\ee We stress that the non-vanishing binorms in Eq.\gl{bior} support 
the completeness of this basis, {\it i.e.} the resolution of identity, \be \qquad\delta(x-x')=\sum\limits_n 
\psi_n(x)\psi_n(x') . \label{decom} \ee Indeed if some of the states in Eq. \gl{decom} were 
"self-orthogonal"  (as it has been accepted in \cite{mois1}) , {\it i.e.} had zero binorms in \gl{bior}, 
the would-be unity in \gl{decom} would annihilate such states thereby signaling the incompleteness.

The PT-symmetry of a potential entails a related symmetry of eigenfunctions, \be V^*(x) = V(-x) \ 
\Longrightarrow \ \tilde\psi_n(x)= \psi^*_n(x)\equiv \gamma_n \psi_n(-x), \quad |\gamma_n| = 1 . \ee when 
the  PT-symmetry is not spontaneously broken, {\it i.e.} $\lambda_n^* = \lambda_n$ (further on, for 
clarity,  we restrict ourselves only with a case of the unbroken PT-symmetry although nearly all results 
can be generalized to the case
 of spontaneous  PT-symmetry breaking with pairs of
 eigenstates having mutually complex conjugated eigenvalues).
The normalization \gl{normal} leads to the value $\gamma_n = \pm 1$ .  One can see that the biorthogonality 
does not, in general, provide positive binorms of states related by the PT-symmetry,

\be \int\limits_{-\infty}^{+\infty}\psi^*_n(-x) \psi_m(x)\,dx=\gamma_n \delta_{nm} = \pm \delta_{nm} , \ee 
bringing negative norm states in the PT-odd sector.

For non-Hermitian Hamiltonians one can formulate the extended eigenvalue problem, searching not only for 
normalizable eigenfunctions but also for normalizable associated functions for discrete part of the energy 
spectrum. Some related problems have been known for a long time in mathematics of linear differential 
equations
 (see for
instance, \cite{naim}) .

Let us give  the formal definition.\\

{\bf Definition 1.} \quad The function $\psi_{n,i}(x)$ is called a formal associated function of $i$-th 
order of the Hamiltonian $h$ for a spectral value $\lambda_n$, if \be 
(h-\lambda_n)^{i+1}\psi_{n,i}\equiv0,\qquad (h-\lambda_n)^{i}\psi_{n,i}\not\equiv 0, \label{canbas1}\ee 
where 'formal' emphasizes that a related function is not necessarily normalizable.
\bigskip

In particular, the associated function of zero order $\psi_{n,0}$ is a formal eigenfunction of $h$ (a 
solution of the homogeneous Schr\"odinger equation, not necessarily normalizable).

Let us single out normalizable associated functions and the case when $h$ maps them into  normalizable 
functions \footnote{It takes place for a certain class of potentials described in Sec. 6, see Part II of 
our paper.}.

Evidently this may occur only for non-Hermitian Hamiltonians. Then for any normalizable associated 
functions $\psi_{n,i}(x)$ and $\psi_{n',i'} (x)$ the transposition symmetry holds 
\be\int\limits_{-\infty}^{+\infty}h \psi_{n,i}(x) \psi_{n',i'}(x)\,dx=\int\limits_{-\infty}^{+\infty} 
\psi_{n,i}(x) h \psi_{n',i'}(x)\,dx . \ee Furthermore one can prove the following relations: \be 
\int\limits_{-\infty}^{+\infty} \psi_{n,i}(x) \psi_{n',i'}(x) \,dx\equiv(\psi^*_{n,i} \ , \psi_{n',i'}) =0, 
\qquad\lambda_n\ne\lambda_{n'},\ee where $(\ldots,\ldots)$ is scalar product.

As well, let's take two normalizable associated functions $\psi_{n,k}(x)$ and $\psi_{n,k'}(x)$ so that, in 
general, $k\not= k' $ and there are two different sequences of associated functions for $i\leq k$ and 
$i'\leq k'$ \be \psi_{n,i}(x)= (h-\lambda_n)^{k-i} \psi_{n,k}(x), \quad \psi_{n,i'}(x)= 
(h-\lambda_n)^{k'-i'} \psi_{n,k'}(x) .\label{canbas2}\ee Then \be  \int\limits_{-\infty}^{+\infty} 
\psi_{n,i}(x) \psi_{n,i'}(x)\,dx=(\psi^*_{n,i} \ , \psi_{n,i'}) =0 ,\qquad  i+i'\le\max\{k,k'\}-1. 
\label{binorm} \ee In particular, for some normalizable associated function $\psi_{n,l}(x)$, the 
"self-orthogonality" \cite{mois1} is realized , \be
 \int\limits_{-\infty}^{+\infty}\psi^2_{n,l}(x) \,dx=0, \qquad
\psi_{n,l}(x) =(h-\lambda)^{i-l}\psi_{n,i}(x), \qquad l=0,\ldots, \Big[{{i-1}\over2}\Big].\ee All the above 
relations are derived from the symmetry of a Hamiltonian under transposition and the very definition of 
associate functions.

We proceed to  the special class of Hamiltonians  for which
 the spectrum is discrete and there is a complete
biorthogonal system
 $\{|\psi_{n,a,i}\rangle,|\tilde\psi_{n,a,i}\rangle\}$ such
that, \ba  && h|\psi_{n,a,0}\rangle=\lambda_n|\psi_{n,a,0}\rangle, \qquad(h-\lambda_n)|\psi_{n,a,i}\rangle
=|\psi_{n,a,i-1}\rangle , \label{schr1}\\
 && h^\dagger|\tilde\psi_{n,a,p_ {n,a}-1}\rangle=\lambda^*_n|\tilde\psi_{n,a,p_{n,a}-1}\rangle,
\quad(h^\dagger-\lambda^*_n)|\tilde\psi_{n,a,p_{n,a}- i -1}\rangle =|\tilde\psi_{n,a,p_{n,a} -i}\rangle, 
\nonumber \ea where
$n=0, 1, 2, \ldots$ is an index of an $h$ eigenvalue $\lambda_n,$\\
$a=1$, \dots, $d_n$ is an index of a Jordan cell (block) for the given eigenvalue,
$\lambda_n$;\\ $d_n$ is a number of Jordan cells for $\lambda_n$;\\
$i=0$, \dots, $p_{n,a}-1$ is an index of associated function in the Jordan cell
with indexes $n,a$\\
and $p_{n,a}$ is a dimension of this Jordan cell. We have taken  a general framework which is applicable 
also for matrix and/or multidimensional Hamiltonians. But the main results of this and the next sections 
are guaranteed only for scalar one-dimensional Hamiltonians with local potentials.

We remark that the number $d_n$ is called a geometric multiplicity of the eigenvalue $\lambda_n$ . For a 
scalar one-dimensional Schr\"odinger equation it cannot normally exceed 1 (but  may reach 2 in specific 
cases of periodic potentials and of potentials unbounded from below).  In turn,  the sum $\sum_a p_{n,a}$ 
is called an algebraic multiplicity of the eigenvalue $\lambda_n$.

The completeness implies the biorthogonality relations 
\be\langle\tilde\psi_{n,a,i}|\psi_{m,b,j}\rangle=\delta_{nm}\delta_{ab} \delta_{ij} \ , \label{assorth}\ee 
and the resolution of identity \be I=\sum\limits_{n=0}^{+\infty}\sum\limits_{a=1}^{d_n}\sum\limits_{i=0 
}^{p_{n,a}-1}|\psi_{n,a,i}\rangle\langle\tilde\psi_{n,a,i}| . \label{resol} \ee The spectral decomposition 
for the Hamiltonian can be constructed as well, \be 
h=\sum\limits_{n=0}^{+\infty}\sum\limits_{a=1}^{d_n}\Big[\lambda_n \sum\limits_{i=0 
}^{p_{n,a}-1}|\psi_{n,a,i}\rangle\langle\tilde\psi_{n,a,i}|+ \sum\limits_{i=0 
}^{p_{n,a}-2}|\psi_{n,a,i}\rangle\langle\tilde\psi_{n,a,i+1}|\Big].\ee It represents the analog of the 
block-diagonal Jordan form for arbitrary non-Hermitian matrices \cite{gant}.

If existing such biorthogonal systems are not unique. Indeed the relations \gl{schr1} remain invariant 
under the group of triangle transformations, \ba
&& |\psi'_{n,a,i}\rangle = \sum\limits_{0\leq j \leq i} \alpha_{ij}|\psi_{n,a,j}\rangle,\nonumber\\
&&|\tilde\psi'_{n,a,k}\rangle =\sum\limits_{k\leq l\leq p_{n,a}-1} \beta_{kl}|\tilde\psi_{n,a,l}\rangle, 
\label{trian} \ea where the matrix elements must obey the following equations, \ba && \alpha_{ij}= 
\alpha_{i+1,\ j+1} = \alpha_{i-j,\ 0}\equiv \alpha_{i-j} ,
\quad \alpha_{00}\not= 0 , \nonumber\\
&&\beta_{kl}= \beta_{k+1,\ l+1}=\beta_{k-l+p_{n,a}-1 \equiv \beta_{k-l},\ p_{n,a}-1} ,\quad 
\beta_{p_{n,a}-1,\ p_{n,a}-1}\not= 0 . \label{alpha}\ea The biorthogonality \gl{assorth} restricts the 
choice of pairs of matrices $\hat\alpha$ and $\hat\beta$  in \gl{trian} to be, \be \hat\beta^\dagger = 
\hat\alpha^{-1} . \ee This freedom in the redefinition of the biorthogonal basis is similar to Eq. 
\gl{normal} and it can be exploited to define the pairs of biorthogonal functions $\psi_{n,a,i} (x) \equiv 
\langle x |\psi_{n,a,i}\rangle$ and $\tilde\psi_{n,a,i} (x) \equiv \langle x|\tilde\psi_{n,a,i}\rangle$ in 
accordance with \gl{bior}.
 However
one has to take into account our enumeration of associated functions $\psi_{n,a,i} (x)$ vs. their 
conjugated ones $\tilde\psi_{n,a,i} (x)$ as it is introduced in Eqs. \gl{schr1} \be \psi_{n,a,i} (x) = 
\tilde\psi^*_{n,a,p_{n,a}-i-1} (x) \equiv \langle\tilde \psi_{n,a,p_{n,a}-i-1}|x\rangle . \label{special} 
\ee Then the analog of  Eq. \gl{bior} reads, \be \int\limits_{-\infty}^{+\infty} \psi_{n,a,i} (x) 
\psi_{m,b,p_{m,b}-j-1} (x) dx = \delta_{nm}\delta_{ab} \delta_{ij}\ \label{bior1}  . \ee We stress that 
this kind of biorthogonal systems is determined uniquely  up to an overall sign.

In these terms it becomes clear that the relations \gl{binorm} have the meaning of orthogonality of some 
off-diagonal pairs in the biorthogonal system $\{|\psi_{n,a,i}\rangle,|\tilde\psi_{n,a,j}\rangle\}$ as
$$
\psi_{n,a,i} (x) = (h - \lambda_n)^{p_{n,a}-1 -i} \psi_{n,a, p_{n,a}-1} (x),$$ \be \tilde\psi^*_{n,a,j} (x)  
= \psi_{n,a, p_{n,a}-1-j} (x) = (h - \lambda_n)^{j} \psi_{n,a, p_{n,a}-1} (x). \label{chain} \ee When 
comparing with specification of indices in Eq. \gl{binorm} one identifies\\ $p_{n,a}-1-j \leftrightarrow 
i,\quad i \leftrightarrow i' $. In both cases $k= k' = p_{n,a}-1$ . Then the inequality \gl{binorm} singles 
out
 off-diagonal binorms, $i \leq j-1$.  From Eq. \gl{chain} it follows
that in order to have all diagonal binorms non-vanishing it is sufficient to prove  that at least one of 
them is not zero because \ba &&\int\limits_{-\infty}^{+\infty} \psi_{n,a,0}(x) \psi_{n,a, p_{n,a}-1}(x)\,dx 
\no &&= \int\limits_{-\infty}^{+\infty} \Big[(h - \lambda_n)^{p_{n,a}-1} \psi_{n,a, p_{n,a}-1} (x)\Big] 
\psi_{n,a, p_{n,a}-1}(x)\,dx \no &&= \int\limits_{-\infty}^{+\infty} \psi_{n,a,i}(x) \psi_{n,a, 
p_{n,a}-1-i}(x)\,dx\not= 0 . \label{normal1} \ea The latter is necessary for the completeness of the basis 
because of the absence of biorthogonal pairs of basis elements made of bound and associated functions in 
the diagonal resolution of identity. If some of such pairs in resolution of identity \gl{resol} were 
biorthogonal then this operator would be at best a projector but not an identity.

For a scalar one-dimensional Schr\"odinger equation the geometric multiplicity $d_n$ of the eigenvalue 
$\lambda_n$
 cannot normally exceed 1 . The latter possibility of non-degenerate eigenstates will be implied throughout 
this paper. Thereby in the rest of the paper the index $a = 1 = d_n$ will be omitted .

It is certainly interesting to examine what are specific features of biorthogonal bases for PT-symmetric 
systems.
   Let us
restrict ourselves by one-dimensional systems with nondegenerate spectrum of eigenstates  and with the
  unbroken PT-symmetry, $\lambda^*_n = \lambda_n$.
First one can easily find that the functions $\tilde\psi_{n,p_{n}-i-1} (-x)$ are normalizable solutions of 
the initial Hamiltonian $h$ and therefore can be decomposed into a linear combination of its basis, 
\be \tilde\psi_{n,p_{n}-i-1} (-x) =  \psi^*_{n,i} (-x) = \sum\limits_{0\leq j \leq i} 
\gamma_{n,ij}\psi_{n,j} (x) , \label{ptbasis} \ee 
where $\gamma_{n,ij} =\gamma_{n,i-j}$ due to Eqs. conjugated to \gl{schr1} . By complex conjugation of 
Eq.\gl{ptbasis} and its further convolution with $\gamma_{n,l-i}$ (and after changing the sign $x 
\rightarrow -x$) one comes to the conditions on matrix elements, \be \sum\limits_{j\leq i \leq l} 
\gamma_{n,l-i} \gamma^*_{n,i-j}\Big |_{j < l} = 0;\quad \gamma_{n,0}^2 =1 . \label{ptcond} \ee The analysis 
of the biorthogonality relations \gl{assorth} and \gl{bior1} for the PT-symmetric basis \gl{ptbasis} leads 
to the conclusion that all numbers $\gamma_{n,i-j}$ are {it real} . Then one derives from \gl{ptcond} that 
$\gamma_{n,i-j}\Big |_{j < i} = 0 $ and the elements of a Jordan cell basis, the eigen- and associated 
functions are simultaneously PT-even or PT-odd depending on the sign of $\gamma_{n,0} = \pm 1$ , \be 
\tilde\psi_{n,p_{n}-i-1} (x) =  \psi^*_{n,i} (x) = \gamma_{n,0}\psi_{n,i} (-x) . \label{ptbasis+} \ee Thus 
a biorthogonal basis of eigen-and associated  functions exists which is fully compatible with PT-symmetry . 
As a consequence of \gl{ptbasis+} the integrals $\int_{-\infty}^{+\infty} \psi_{n,0}(x) \psi^*_{n, 
p_{n}-1}(-x)\,dx$ remain real.

We remark that in general case the existence and the completeness of a  biorthogonal system  is not obvious 
(especially if the continuum spectrum is present \cite{sokancan}) and needs a careful examination.

\section{The origin of non-diagonalizable Hamiltonians is level confluence}

Let us demonstrate that one can generate a Jordan cell of a Hamiltonian in the process of coalescing levels 
of the Hamiltonian . In the simplest case one can consider a Hamiltonian depending on the parameter $\mu$,
$$h_\mu=-\partial^2+V(x;\mu)$$
with two eigenfunctions $\psi_{1,2}(x;\mu)$,
$$h_\mu\psi_{1,2}(x;\mu)=\lambda_{1,2}(\mu)\psi_{1,2}(x;\mu).$$
Assume that the levels $\lambda_1(\mu)$ and $\lambda_2(\mu)$ coalesce for $\mu=\mu_0$:
$$\psi_1(x;\mu_0)\equiv\psi_2(x;\mu_0)=\varphi_0(x),\qquad\lambda_1(\mu_0)=
\lambda_2(\mu_0)=\lambda_0.$$ Let us also suppose that the functions ${{\partial\psi_1}\over{\partial\mu}} 
(x;\mu)$ and ${{\partial\psi_2}\over{\partial\mu}}(x;\mu)$ are normalizable. Then, it is evident that
$$[h_\mu-\lambda_{1,2}(\mu)]{{\partial\psi_{1,2}}\over{\partial\mu}}
=-\Big[{{\partial V}\over{\partial\mu}}-\lambda'_{1,2}(\mu)\Big]\psi_{1,2},$$
$$(h_{\mu_0}-\lambda_0)\Big[{{\partial\psi_1}\over{\partial\mu}}(x;\mu_0)
-{{\partial\psi_2}\over{\partial\mu}}(x;\mu_0)\Big]=[\lambda'_1(\mu_0)- \lambda'_2(\mu_0)]\varphi_0(x)$$ 
and, thus, the functions
\begin{equation}\varphi_0(x),\qquad \varphi_1(x)={{{{\partial\psi_1}\over{\partial\mu}}
(x;\mu_0)-{{\partial\psi_2}\over{\partial\mu}}(x;\mu_0)}\over{\lambda'_1 
(\mu_0)-\lambda'_2(\mu_0)}}\la{phi1}\end{equation} form a Jordan cell of the second order for the  
Hamiltonian $h_0=h_{\mu_0}$:
$$h_0\varphi_0=\lambda_0\varphi_0,\qquad(h_0-\lambda_0)\varphi_1=\varphi_0.$$

Let us proceed now to the case with three coalescing levels of $h_\mu$:
$$[h_\mu-\lambda_j(\mu)]\psi_j(x;\mu)=0,\qquad j=1,2,3,$$
$$\psi_1(x;\mu_0)\equiv\psi_2(x;\mu_0)\equiv\psi_3(x;\mu_0)=\varphi_0(x),
\qquad\lambda_1(\mu_0)=\lambda_2(\mu_0)=\lambda_3(\mu_0)=\lambda_0.$$ Let us introduce the  auxiliary 
functions
$$\psi_1^{(0)}(x;\mu)=[1+\varkappa(\mu-\mu_0)]\psi_1(x;\mu),\qquad
\psi_j^{(0)}(x;\mu)=\psi_j(x;\mu),\quad j=2,3,$$
$$\psi_j^{(1)}(x;\mu)={{{{\partial\psi_j^{(0)}}\over{\partial\mu}}(x;\mu)-
{{\partial\psi_3^{(0)}}\over{\partial\mu}}(x;\mu)}\over{\lambda'_j(\mu)- \lambda'_3(\mu)}},\qquad j=1,2,$$ 
where the constant $\varkappa$ is chosen so that the associated functions of the first order
$$\psi_1^{(1)}(x;\mu_0)={{{{\partial\psi_1}\over{\partial\mu}}(x;\mu_0)-
{{\partial\psi_3}\over{\partial\mu}}(x;\mu_0)}\over{\lambda'_1(\mu_0)- 
\lambda'_3(\mu_0)}}+{\varkappa\over{\lambda'_1(\mu_0)- \lambda'_3(\mu_0)}}\varphi_0(x)$$ and
$$\psi_2^{(1)}(x;\mu_0)={{{{\partial\psi_2}\over{\partial\mu}}(x;\mu_0)-
{{\partial\psi_3}\over{\partial\mu}}(x;\mu_0)}\over{\lambda'_2(\mu_0)- \lambda'_3(\mu_0)}}$$ (cf. with 
\gl{phi1}) are identical\footnote{The constant $\varkappa$ exists because in one-dimensional case the 
difference of normalizable associated functions of the first order for the same eigenfunction is 
proportional to this eigenfunction.This freedom has been discussed in the previous section, see 
Eq.\gl{trian} .}. When using these auxiliary functions one can obtain the canonical set of associated 
functions,
$$\varphi_0(x),\qquad\varphi_1(x)=\psi_1^{(1)}(x;\mu_0)\equiv
\psi_2^{(1)}(x;\mu_0),\qquad\varphi_2(x)={{{{\partial\psi_1^{(1)}} 
\over{\partial\mu}}(x;\mu_0)-{{\partial\psi_2^{(1)}}\over{\partial\mu}} 
(x;\mu_0)}\over{2[\lambda'_1(\mu_0)-\lambda'_2(\mu_0)]}},$$ which form Jordan cell of the third order for 
the Hamiltonian $h_0=h_{\mu_0}$:
$$h_0\varphi_0=\lambda_0\varphi_0,\qquad(h_0-\lambda_0)\varphi_j=\varphi_{j-1},
\quad j=1,2.$$ Thus, we have shown that the confluence of two (three) levels (of algebraic multiplicity 1) 
leads to appearance of a Jordan cell of the second (third) order. The described construction is illustrated 
by an example in Subsec. 7.2. In the case of confluence of a larger number $n$ of levels (of algebraic 
multiplicity 1) one can construct the canonical chain of $n$ normalizable eigenfunction and associated 
functions in the same way.

One can examine also the confluence of levels of different algebraic multiplicity. Let us restrict 
ourselves to the simplest case, when for the Hamiltonian $h_\mu$ level $\lambda_1(\mu)$ of algebraic 
multiplicity 2 coalesces for $\mu=\mu_0$ with level $\lambda_2(\mu)$ of algebraic multiplicity 1:
$$h_\mu\psi_{10}=\lambda_1(\mu)\psi_{10},\qquad [h_\mu-\lambda_1(\mu)]
\psi_{11}=\psi_{10},$$
$$h_\mu\psi_2=\lambda_2(\mu)\psi_2,$$
$$\psi_{10}(x;\mu_0)\equiv\psi_2(x;\mu_0)=\varphi_0(x),\qquad\lambda_1(\mu_0)=
\lambda_2(\mu_0)=\lambda_0.$$ Again we introduce the  auxiliary functions
$$\psi_1^{(0)}(x;\mu)=\psi_{10}(x;\mu),\qquad\psi_2^{(0)}(x;\mu)=[1+\varkappa
(\mu-\mu_0)]\psi_2(x;\mu),$$
$$\psi_1^{(1)}(x;\mu)=\psi_{11}(x;\mu),\qquad\psi_2^{(1)}(x;\mu)=
{{{{\partial\psi_1^{(0)}}\over{\partial\mu}}(x;\mu)-{{\partial\psi_2^{(0)}} 
\over{\partial\mu}}(x;\mu)}\over{\lambda'_1(\mu)-\lambda'_2(\mu)}},$$ where one should choose $\varkappa$ 
so that the associated functions of the first order \\$\psi_1^{(1)}(x;\mu_0)\equiv\psi_{11}(x;\mu_0)$ and
$$\psi_2^{(1)}(x;\mu_0)\equiv
{{{{\partial\psi_{10}}\over{\partial\mu}}(x;\mu_0)-{{\partial\psi_2} 
\over{\partial\mu}}(x;\mu_0)}\over{\lambda'_1(\mu_0)-\lambda'_2(\mu_0)}}- 
{\varkappa\over{\lambda'_1(\mu_0)-\lambda'_2(\mu_0)}}\varphi_0(x), $$ are identical. Therefrom one can get 
the appropriate set of associated functions 
\ba&&\varphi_0(x),\qquad\varphi_1(x)=\psi_1^{(1)}(x;\mu_0)\equiv 
\psi_2^{(1)}(x;\mu_0),\nonumber\\&&\varphi_2(x)={{{{\partial\psi_2^{(1)}} 
\over{\partial\mu}}(x;\mu_0)-2{{\partial\psi_1^{(1)}}\over{\partial\mu}} 
(x;\mu_0)}\over{2[\lambda'_2(\mu_0)-\lambda'_1(\mu_0)]}},\ea which form Jordan cell of the third order for 
the Hamiltonian $h_0=h_{\mu_0}$:
$$h_0\varphi_0=\lambda_0\varphi_0,\qquad(h_0-\lambda_0)\varphi_j=\varphi_{j-1},
\quad j=1,2.$$

\section{Non-linear SUSY for complex potentials}
Supersymmetric  Quantum Mechanics (SUSY QM) in one dimension represents a concise way for an almost 
isospectral transformation between two quantum systems \cite{nico} -- \cite{sukum} (see the reviews 
\cite{genden} -- \cite{lima} ). Conventionally it can be built for a pair of Hamiltonians $h^+$ and $h^-$ 
assembled
 into a Super-Hamiltonian,
\ba H = \left(\begin{array}{cc}
h^+& 0\\
0 & h^-
\end{array}\right) =\left(\begin{array}{cc}
-\partial^2 + V_1(x)& 0\\
0 & - \partial^2 + V_2(x)
\end{array}\right) \equiv -\partial^2 {\bf I} + {\bf V}(x), \la{suham}
\ea where the potential ${\bf V}(x) $ is, in general, complex. The (almost) isospectral connection between  
$h^+$ and $h^-$
 is realized by  the intertwining relations ,
\be h^+ q^+ = q^+ h^-, \quad q^- h^+ = h^- q^-, \la{intertw} \ee with $q^\pm$ being components of the 
supercharges, \ba Q=\left(\begin{array}{cc}
 0 &  q^+\\
0 & 0
\end{array}\right),\quad
\bar Q=\left(\begin{array}{cc}
 0 &  0\\
q^- & 0
\end{array}\right),\quad  Q^2 = \bar Q^2 = 0. \la{such}
\ea The isospectral relations \gl{intertw}  result in the conservation of supercharges or the supersymmetry 
of the Super-Hamiltonian, \ba [H,Q] = [H,\bar Q] =0, \la{cons} \ea

In general, its algebraic closure is given, by  a non-linear (deformed) SUSY algebra, \be \left\{Q,\bar 
Q\right\} = {\cal P} (H), \la{poly} \ee where ${\cal P} (H)$ is a function of the Super-Hamiltonian 
\cite{acdi,ancan,ansok}.

The relevant supercharges are supposed to be generated by $N$-th order differential operators with smooth 
coefficient functions $ w^{\pm}_k (x)$: \be q^{\pm} \equiv q^{\pm}_N =\sum_{k=0}^N w^{\pm}_k (x)\partial^k, 
\quad w^{\pm}_N = const \equiv (\mp 1)^N. \la{superch} \ee We focus our analysis on the non-Hermitian 
Hamiltonians interrelated by complex supercharges, {\it i.e.} the supercharges with complex and smooth 
coefficient functions $w^{\pm}_k (x)$ . In this paper we choose $q^-_N$ connected to $q^+_N$ by means of 
$^t$ -- transposition, $q^-_N = (q^+_N)^t$ . Evidently the Super-Hamiltonian of Schr\"odinger type is 
self-transposed ($^t$-symmetric) as only the scalar potentials are under consideration.

The algebraic structure of a Non-linear SUSY for local Hamiltonians is exhaustively determined
by the following theorem,\\

\noindent
{\bf Theorem 1}: \underline{on SUSY algebras with transposition symmetry}\\
{\it
Let us introduce two sets of $N$ linearly independent functions\\
$\phi^{\pm}_{n}(x)$ $(n=1,\cdots N)$  which represent complete sets of zero-modes of the supercharge 
components,
\begin{eqnarray}
q^{\pm}_N \phi^{\pm}_{n}=0,\quad q^-_N =(q^+_N)^t. \la{tmode}
\end{eqnarray}
Then:\\
\noindent 1) the Hamiltonians $h^{\pm}$ have finite matrix representations when acting on the set of 
functions $\phi^{\pm}_{n}(x)$,
\begin{eqnarray}
h^{\pm}\phi^{\mp}_{n}=\sum_{m} S^{\pm}_{nm}\phi^{\mp}_{m}, \la{quasiham1}
\end{eqnarray}

\noindent 2) the SUSY algebra closure with $\bar Q = Q^t$ takes the polynomial form,
\begin{eqnarray}
\left\{Q, Q^t\right\} = {\rm det} \left[E{\bf I}-{\bf S}^{+}\right]_{E = H} ={\rm det} \left[E{\bf I}-{\bf 
S}^{-}\right]_{E = H} \equiv {\cal P}_N (H).\la{tposusy}
\end{eqnarray}
}\\
The proof in \cite{ansok} is based on the quasi-diagonalization of matrices ${\bf S}^{\pm}$, {\it i.e.} on 
their reduction to the Jordan canonical form  $\widetilde{\bf S}^{\pm}$ which is block-diagonal. Such a 
diagonalization can be realized by non-degenerate linear transformations $\Omega^\pm$ of the zero-mode sets 
which induce the similarity transformations of matrices ${\bf S}^{\pm}$, \ba &&\tilde\phi^\pm_l = 
\sum_{m=1}^N \Omega^\mp_{lm} \phi^\pm_m,\quad h^\pm \tilde\phi^\mp_l =  \sum_{m=1}^N 
\widetilde{S}^{\pm}_{lm} \tilde\phi^\mp_m,\qquad \no &&\widetilde{\bf S}^{\pm} = \Omega^\pm {\bf S}^\pm 
\left(\Omega^\pm\right)^{-1}. \la{jord} \ea Evidently the canonical bases of zero-modes of intertwining 
operators $q_N^\pm$ form the set of (in general, formal -- not necessarily normalizable) solutions and 
associated functions of the Hamiltonians $h^\mp$. These elements of canonical bases of intertwining 
operator kernels are named as {\it transformation functions}. The proof in \cite{ansok} can be easily 
generalized to the complex case with transposition symmetry being built up along the same scheme.

For the polynomial SUSY algebra there is a possibility that the intertwining operators may be trivially 
reduced by a factor depending on the Hamiltonian, without any changes in the Hamiltonians $h^\pm$, namely, 
\be q_N^\pm=P(h^\pm)p_M^\pm=p_M^\pm P(h^\mp) , \la{triv} \ee where $P(x)$ is assumed to be a polynomial and 
$N \geq M +2$. Thus some of the roots of associated polynomials may not be involved in determination of the 
structure of the potentials.

This problem of disentangling the nontrivial part of a supercharge and avoiding multiple SUSY algebras 
generated by means of ``dressing''
 can be systematically tackled with the help of the
following theorem which can be also extended for complex potentials.\\

\noindent
\underline{``Strip-off'' {\bf Theorem 2 .}}\\
{\it Let's assume the construction of the Theorem on SUSY algebras with transposition symmetry. Then the
requirement\\
that the matrix $\widetilde{\bf S}^-$ (or $\widetilde{\bf S}^+$) generated on the subspace of zero-modes of 
the operator $q^+_N$ (or $q^-_N$) contains $m$ pairs (and no more) of Jordan cells with equal eigenvalues  
$\lambda_l$ in each pair
and the sizes $\delta k_l$ and $k_l+\delta k_l$ ($\delta k_l$ being the size of a smallest cell in the 
$l$-th pair)\\
is necessary and sufficient to ensure that the intertwining operator $q^+_N$ (or $q^-_N$) can be 
factorized: \be q^\pm_{N} =  p^\pm_{M} \prod^m_{l=1} (\lambda_l - h^\mp)^{\delta k_l}, \la{factor} \ee 
where $ p^\pm_{M}$ are intertwining operators of order $M= N - 2 \sum^m_{l=1} \delta k_l = \sum^n_{j=1}  
k_j $ which cannot be  decomposed
further on in the product \gl{factor} type. Herein $k_j$ for $ m+1\le j\le n$ are sizes of unpaired Jordan 
cells .}\\

\nn \underline{Remark 1.} The matrices  $\widetilde{\bf S}^\pm$ cannot contain more than {\bf two} Jordan 
cells with the {\bf same} eigenvalue $\lambda$ because otherwise the operator $\lambda - h^\pm$ would have 
more than two linearly independent
zero-mode solutions.\\

\nn \underline{Remark 2.}  This theorem together with the Theorem 1 entails the essential identity of the 
Jordan forms
 $\widetilde{\bf S}^-$ and  $\widetilde{\bf S}^+$ (up to transposition of
certain Jordan cells).\\

\nn \underline{Remark 3.} The supercharge components cannot be stripped-off if the polynomial ${\cal P}_N 
(x)$ does not have degenerate zeroes. The latter is sufficient to deal with  SUSY charges non-trivially 
factorizable, but not necessary because  degenerate zeroes may well arise in the ladder (dressing chain) 
construction giving
new pairs of isospectral potentials.\\

\nn \underline{Remark 4.} In general, a Super-Hamiltonian may commute with several different supercharges. 
In this case few hidden-symmetry operators arise. The optimization of such a system of supercharges till 
one or two independent ones, their stripping-off and the minimal structure of a symmetry operator has been 
investigated in \cite{ansok} in details for Hermitian Hamiltonians. In so far as in the quoted paper the 
transposition was used  as a main conjugation operation,  all essential results of \cite{ansok} remain to 
be valid also for
complex potentials.\\

The intertwining operators (supercharge components) can be formally factorized into the products of 
elementary Darboux operators. Let $\phi^-_j \equiv \phi_j$, $j=1$, \dots, $N$ be\footnote{In what follows, 
in the notations for $\phi^\mp_j$, we omit their relation to  a Hamiltonian $h^\pm$ when it is only one of 
these sets which is used, in order to avoid too heavy indices. } the basis, in which the $\bf S^+$-matrix 
(see Th.~1) has a canonical Jordan form and $\lambda_j$ is an eigenvalue of $\bf S^+$ corresponding to the 
Jordan cell, to  which $\phi_j$ belongs. Then adapting the Lemma 1 from \cite{ansok} for non-Hermitian 
Hamiltonians one can prove the following statements:

1) for the supercharge components the factorization holds, in particular, \be q_N^-=r_N^-\ldots r_1^-, \ee 
where the Darboux operators
$$r_j^-=\partial+\chi_j(x),\qquad j=1,\ldots,N,$$
 can be chosen so that
\begin{equation}r_j^-\ldots r_1^-\phi_{N-j+1}=0,\qquad j=1,\ldots,N;
\la{alpha1}\end{equation}

2) the chain relations take place
$$ (r_{l+1}^-)^tr_{l+1}^-+\lambda_{l+1}=r_{l}^-(r_l^-)^t+\lambda_{l}\equiv
h_l,\qquad j=1,\ldots,N-1,$$ \be (r_1^-)^tr_1^-+\lambda_1=h^+\equiv h_0,\qquad 
r_N^-(r_N^-)^t+\lambda_N=h^-\equiv h_N;\ee

3) the intermediate Hamiltonians $h_l$, $l=1$, \dots, $N-1$ have Schr\"odinger form:
$$h_l=-\partial^2+v_l(x),\qquad
v_l(x)=\chi^2_{l+1}(x)-\chi'_{l+1}(x)+\lambda_{l+1}=\chi^2_{l}(x)+\chi'_{l}(x)+\lambda_{l},$$
\begin{equation} V_1(x)\equiv
v_0(x)=\chi^2_{1}(x)-\chi'_{1}(x)+\lambda_{1},\qquad V_2(x)\equiv v_N(x)= 
\chi^2_{N}(x)+\chi'_{N}(x)+\lambda_{N}, \la{beta}\end{equation} but, in general, with complex and/or 
singular potentials;

4) the intertwining relations are valid: \be h_l(r_{l+1}^-)^t=(r_{l+1}^-)^th_{l+1},\qquad 
r_{l+1}^-h_l=h_{l+1} r_{l+1}^-,\qquad l=0,\ldots,N-1. \ee

Let us introduce the generalized Crum determinants made of solutions of the initial Schr\"odinger equation 
for the Hamiltonian $h^+$ {\it as well as} of some of its formal associated functions,
\be w_j(x)=\begin{vmatrix}\phi_N(x)&\phi'_N(x)&\ldots&\phi_N^{(j-1)}(x)\\
\phi_{N-1}(x)&\phi'_{N-1}(x)&\ldots&\phi_{N-1}^{(j-1)}(x)\\ \hdotsfor{4}\\
\phi_{N-j+1}(x)&\phi'_{N-j+1}(x)&\ldots&\phi_{N-j+1}^{(j-1)}(x)\\
\end{vmatrix},\qquad j=1,\ldots,N. \ee
Then in virtue of Eq. \gl{alpha1} one finds the representation for the intertwining operators, \be 
r_j^-\ldots r_1^-={1\over w_j(x)}
\begin{vmatrix}\phi_N(x)&\phi'_N(x)&\ldots&\phi_N^{(j)}(x)\\
\hdotsfor{4}\\
\phi_{N-j+1}(x)&\phi'_{N-j+1}(x)&\ldots&\phi_{N-j+1}^{(j-1)}(x)\\
1&\partial&\ldots&\partial^j\end{vmatrix},\qquad j=1,\ldots,N \ee and, consequently, \be r_j^-\ldots 
r_1^-\phi_{N-j}={{w_{j+1}(x)}\over{w_j(x)}},\qquad j=1, \ldots,N-1. \ee Hence the intermediate 
superpotentials are uniquely determined by the chosen basis of solutions and formal associated functions of 
a given Hamiltonian $h^+$ for a given ordering , \be 
\chi_j(x)=-{{[w_j(x)/w_{j-1}(x)]'}\over{w_j(x)/w_{j-1}(x)}}= 
-{{w'_j(x)}\over{w_j(x)}}+{{w'_{j-1}(x)}\over{w_{j-1}(x)}},\qquad j=1, \ldots,N,\quad w_0(x)\equiv1.\ee 
Thus, from Eq. \gl{beta} one obtains the chain relations between intermediate potentials, \be 
v_j(x)-v_{j-1}(x)=2\chi'_j(x)=-2\Big[\ln{{w_j(x)}\over{w_{j-1}(x)}}\Big]'', \qquad j=1, \ldots, N; \ee and 
furthermore \be v_j(x)-v_0(x)=-2[\ln w_j(x)]'',\qquad j=0, \ldots, N. \ee It leads finally to the 
connection between the components of the potential in the Super-Hamiltonian, \be V_2(x)=V_1(x)-2[\ln 
w_N(x)]''.\la{v2v1}\ee The above set of relations is well-known from the Crum theory \cite{matv}. However 
here we have extended them including not only solutions of a Schr\"odinger equation but also its formal 
associated functions.\\

\nn \underline{Remark 5.}  Let's comment the sufficient conditions to preserve PT-symmetry under SUSY 
transformations, {\it i.e.}  after intertwining with a Darboux-Crum operator . They consist in requirement 
for all transformation functions (all elements of the zero-mode basis of the intertwining operator) to be 
symmetric or antisymmetric in respect to PT-reflection . Indeed the PT-(anti)symmetry of basis elements 
entails the PT-(anti)symmetry of their Wronskian wherefrom one derives the PT-antisymmetry of $W'/W$ and 
the PT-symmetry of $V_2=V_1-2(\ln W)''$ . This is why under such conditions the above presented theorems 
and the forthcoming ones are certainly valid .

\section{Non-diagonalizable Hamiltonians and normalizability of associated functions}
\hspace*{2ex} In this section we start examination of how the Darboux transformation may change the 
structure of a non-diagonalizable Hamiltonian with a Jordan cell spanned by a set of associated functions. 
We keep in mind that an intertwining operator may annihilate part of them as zero modes. As a result we 
find the additive composition of Jordan cells for partner Hamiltonians $h^\pm$ mediated by a Jordan cell 
for the Hamiltonian mapping of the zero-mode subspace of the intertwining operator $q^-_N$.

 The rigorous results can be obtained with a specification of
the class of potentials invariant under Darboux transformations. For such a class the asymptotic 
normalizability  of associated functions at one of the infinities is  preserved by Darboux transformations 
and will be described in certain lemmas and corollaries. When the normalizability on the whole axis is 
achieved  we call the related associated functions as normalizable in general. The detailed mathematical 
proofs  are given  in part II,  here  we give  only the general ideas of the construction as well as the 
formulations of the theorems and lemmas and the related corollaries .

Let us investigate the Jordan structure of the Hamiltonians $h^\pm$. In what follows
we restrict ourselves  to the particular  class of potentials:\\

{\bf Definition 2.} Let $K$ be the set of all potentials $V(x)$ such that:

1) $V(x)\in C_{\mathbb R}^\infty$;

2) there are $R_0>0$ and $\varepsilon>0$ ($R_0$ and $\varepsilon$ depend on $V(x)$) such that for any 
$|x|\ge R_0$ the inequality ${\rm{Re}}\,V(x)\ge\varepsilon$ takes place;

3) ${\rm {Im}}\,V(x)/{\rm {Re}}\,V(x)=o(1),$ $x\to\pm\infty$ (this is sufficient to ensure the reality of 
the continuous spectrum);

4) functions \be \bigg(\int\limits_{\pm R_0}^x\sqrt{|V(x_1)|}dx_1\bigg)^2 
\bigg({{|V'(x)|^2}\over{|V(x)|^3}}+{{|V''(x)|}\over{|V(x)|^2}}\bigg) \ee
are bounded respectively for $x\ge R_0$ and $x\le -R_0$.\\

\nn \underline{Remark 6.} The last condition  is not very rigid: it is fulfilled (if $x\to+\infty$), for 
example, for potentials: \hfill \hfill \linebreak \centerline{1)\hfill$V(x)=ax^\gamma[1+o(1)], \qquad 
V'(x)=a\gamma x^{\gamma-1}[1+o(1)],\qquad\qquad$\hfill}
$$V''(x)=a\gamma(\gamma-1)x^{\gamma-2}+o(x^{\gamma-2}),\qquad
a>0,\quad\gamma>0;$$ \centerline{2)\hfill$V(x)=V_0+ax^{-\gamma}[1+o(1)], \qquad V'(x)=-a\gamma 
x^{-\gamma-1}[1+o(1)],\qquad\qquad$\hfill}
$$V''(x)=a\gamma(\gamma+1)x^{-\gamma-2}[1+o(1)],\qquad
V_0>0,\quad a\in\mathbb{C},\quad{\rm{Re}}\,\gamma>0;$$ 
\centerline{3)\hfill$V(x)=ax^{\alpha}e^{bx^\beta}[1+o(1)], \qquad V'(x)=ab\beta 
x^{\alpha+\beta-1}e^{bx^\beta}[1+o(1)],\qquad\qquad$\hfill}
$$V''(x)=ab^2\beta^2x^{\alpha+2\beta-2}e^{bx^\beta}[1+o(1)],\qquad
a>0,\quad b>0,\quad\alpha\in\mathbb R,\quad\beta>0;$$ 
\centerline{4)\hfill$V(x)=V_0+ax^{\alpha}e^{-bx^\beta}[1+o(1)], \qquad V'(x)=-ab\beta 
x^{\alpha+\beta-1}e^{-bx^\beta}[1+o(1)],\qquad\qquad$\hfill}
$$V''(x)=ab^2\beta^2x^{\alpha+2\beta-2}e^{-bx^\beta}[1+o(1)],\qquad
V_0>0,\,a\in\mathbb{C},\,{\rm{Re}}\,b>0,\,\alpha\in\mathbb C,\,\beta>0.$$ A similar statement is valid for 
$x\to-\infty$. Thus one can find in this class of potentials both representatives with
purely discrete spectrum and the Hamiltonians with continuum spectrum .\\

\nn \underline{Remark7.} This class of potentials only partially overlaps with PT-symmetric set of 
potentials investigated recently \cite{bender}-\cite{levai},\cite{dorey}, \cite{jones} . We call the 
related class of Hamiltonians as softly non-Hermitian as they certainly don't involve complex coordinates 
in definition of asymptotic boundary conditions.

The set $K$ is closed under intertwining of Hamiltonians, that
follows from the\\

\underline{{\bf  Lemma 1}: on invariance of the potential set $K$ .} {\it Let: 1) $h^+=-\partial^2+V_1(x)$, 
$V_1(x)\in K$; 2) $h^-=-\partial^2+V_2(x)$, $V_2(x)\in C_{\mathbb R}$; 3) $q_N^-h^+=h^-q_N^-$, where 
$q_N^-$ is differential operator of $N$th order with coefficients from $C_{\mathbb R}^2$; 4) each 
eigenvalue of ${\bf S}^+$-matrix of $q_N^-$ (see Th. 1) satisfies one of the conditions: either 
$\lambda\le0$ or ${\rm Im}\, \lambda\ne0$. Then: 1) $V_2(x)\in K$; 2) coefficients of $q_N^-$ belong to 
$C_{\mathbb R}^\infty$; 3) $h^+q_N^+=q_N^+h^-$, where $q_N^+=(q_N^-)^t$, and moreover coefficients of 
$q_N^+$
belong to $C_{\mathbb R}^\infty$ also.}\\

Let us now analyze the normalizability properties of associated
functions.\\

{\bf Definition 3.} A function $f(x)$ is called {\it normalizable} at $+\infty$ (at $-\infty$), if there is 
$R_+$ ($R_-$) such that \be \int\limits_{R_+}^{+\infty}|f(x)|^2\,dx<+\infty\qquad 
\bigg(\int\limits_{-\infty}^{R_-}|f(x)|^2\,dx<+\infty\bigg).\ee
Otherwise $f(x)$ is called {\it non-normalizable} at $+\infty$ (at $-\infty$).\\

Using the asymptotics of formal associated functions (see lemma 9, part II), one can show that:

1) when the potential in $h$ (by which, in what follows, we imply either $h^+$ or $h^-$) belongs to the 
class $K$, one can show that any formal associated function of $n$-th order of $h$, normalizable at 
$+\infty$ or $-\infty$ respectively, for the spectral value $\lambda$, satisfying either $\lambda \le0$ or 
${\rm{Im}\,} \lambda \ne0$, can be decomposed as follows 
\be\sum\limits_{j=0}^na_{j,\uparrow\downarrow}\varphi_{j,\uparrow\downarrow}(x),\qquad 
a_{j,\uparrow\downarrow}={\rm{Const}},\quad a_{n,\uparrow\downarrow}\ne0,\la{2.9} \ee where 
$\varphi_{j,\uparrow\downarrow}(x)$, $j\ge0$ stand for either $\varphi_{j,\uparrow}(x)$ or 
$\varphi_{j,\downarrow}(x)$ and they form  a sequence of  associated functions normalizable at $+\infty$ or 
$-\infty$ respectively, \be h\varphi_{0,\uparrow\downarrow}=\lambda 
\varphi_{0,\uparrow\downarrow},\qquad(h-\lambda)\varphi_{j,\uparrow\downarrow} 
=\varphi_{j-1,\uparrow\downarrow},\quad j\ge1\ee Correspondingly, any  associated function of $n$-th order 
of $h$, non-normalizable at the same $+\infty$ or $-\infty$, for the same spectral value $\lambda$ can be 
presented as follows \be\sum\limits_{j=0}^n\big(b_{j,\uparrow\downarrow}\varphi_{j,\uparrow\downarrow}(x)+ 
c_{j,\uparrow\downarrow}\hat\varphi_{j,\uparrow\downarrow}(x)\big),\ee where 
$b_{j,\uparrow\downarrow},c_{j,\uparrow\downarrow}={\rm{const}}$, either $b_{n,\uparrow\downarrow}\ne0$ or 
$c_{n,\uparrow\downarrow}\ne0$ and $\hat\varphi_{j,\uparrow\downarrow}(x)$, $j\ge0$ form a sequence of 
non-normalizable at $+\infty$ or $-\infty$ respectively associated functions \be 
h\hat\varphi_{0,\uparrow\downarrow}=\lambda\hat\varphi_{0,\uparrow\downarrow}, 
\qquad(h-\lambda)\hat\varphi_{j,\uparrow\downarrow} =\hat\varphi_{j-1,\uparrow\downarrow},\quad j\ge1;\ee

2) for the Hamiltonian  with a potential from the class $K$,  there are no  degenerate eigenvalues, 
satisfying either $\lambda\le0$ or ${\rm{Im}} \,\lambda\ne0$, {\it i.e.} the eigenvalues, whose geometric 
multiplicity exceeds 1 (the eigenvalues, for which there are more than one linearly independent 
eigenfunctions). Hence, for the Hamiltonian with a potential from $K$ there is no more than one Jordan cell 
made of an eigenfunction and associated functions, normalizable on the whole axis, for any given eigenvalue 
$\lambda$ such that either $\lambda\le0$ or ${\rm{Im}}\,\lambda\ne0$.

Properties of associated functions under intertwining are described
by the\\

{\bf Lemma 2.} {\it Let: 1) the conditions of the lemma 1 take place; 2) $\varphi_n(x)$, $n=0$, \dots $M$ 
be a sequence of associated functions of $h^+$ for the spectral value $\lambda$:
$$h^+\varphi_0=\lambda\varphi_0,\qquad (h^+-\lambda)\varphi_n=
\varphi_{n-1},\quad n\ge1,$$ where either $\lambda\le0$ or ${\rm{Im}}\,\lambda\ne0$. Then:

1) there is a number $m$ such that $0\le m\le\min\{M+1,N\}$,
$$ q_N^-\varphi_n\equiv0, \qquad n<m, $$
and
$$\psi_l=q_N^-\varphi_{m+l},\qquad l=0,\ldots, M-m$$
is a sequence of associated functions of $h^-$ for the spectral value $\lambda$:
$$h^-\psi_0=\lambda\psi_0,\qquad (h^--\lambda)\psi_l=
\psi_{l-1},\quad l\ge1;$$

2) if the function $\varphi_{n}(x)$, for a given $0\le n\le M$, is normalizable at $+\infty$ (on 
$-\infty$), then the function $q_N^-\varphi_{n}$ is normalizable at $+\infty$ (on
$-\infty$) as well.}\\

{\bf Corollary 1.} Since $h^+$ is an intertwining operator for itself and both eigenvalues of its ${\bf 
S^+}$-matrix are zero, then if $\varphi_{n}(x)$ is normalizable at $+\infty$ (at $-\infty$), then 
$\varphi_j(x)$, $j=0$, \dots $n-1$ are normalizable at
$+\infty$ (at $-\infty$) as well.\\

{\bf Corollary 2.} If there is an associated normalizable function  $\varphi_n(x)$ of $n$-th order of the 
Hamiltonian $h$ with a potential from $K$, for an eigenvalue $\lambda$, which is either $\lambda\le0$ or 
${\rm {Im}}\,\lambda\ne0$, then for this eigenvalue there is an associated function $\varphi_j(x)$ of 
Hamiltonian $h$, normalizable on the whole axis, of any smaller order $j$:
$$\varphi_j=(h-\lambda)^{n-j}\varphi_n,\qquad j=0,\ldots,n-1.$$
\\

{\bf Corollary 3.} Let $\varphi^-_{i,j}(x)$ be a canonical basis of zero-modes of the intertwining operator 
$q_N^-$, {\it i.e.} such that ${\bf S^+} $-matrix (from the Theorems 1 and 2) has in this basis the 
canonical (Jordan) form:
$$h^+\varphi^-_{i,0}=\lambda_i\varphi^-_{i,0},\qquad(h^+-\lambda_i)
\varphi^-_{i,j}=\varphi^-_{i,j-1},\quad j=1,\ldots, k_i-1,$$ where $k_i$ is a rank of a Jordan cell for  
$\lambda_i$. Then there are numbers $k^+_{i\uparrow}$ and $k^+_{i\downarrow}$, $0\le 
k^+_{i\uparrow,\downarrow}\le k_i$ related to the Hamiltonian $h^+$ such that for any $i$ the functions
$$\varphi^-_{i,j}(x),\qquad j=0,\ldots, k^+_{i\uparrow,\downarrow}-1$$
are normalizable at $+\infty$ or $-\infty$ respectively and the functions
$$\varphi^-_{i,j}(x),\qquad j=k^+_{i\uparrow,\downarrow},\ldots,k_i-1$$
are non-normalizable at the same $+\infty$ or $-\infty$ .  Thus one can derive that the number of  
functions $\varphi^-_{i,j}(x)$ normalizable on the whole axis is equivalent to $\min\lbrace 
k^+_{i\uparrow}, k^+_{i\downarrow}\rbrace$ and the number of  functions $\varphi^-_{i,j}(x)$ 
non-normalizable at both ends is given by $k_i - \max\lbrace k^+_{i\uparrow}, k^+_{i\downarrow}\rbrace$ .

Independence of these numbers $k^+_{i\uparrow,\downarrow}$ on a choice of the canonical basis, in the case 
of stripped-off $q_N^-$ , follows
from\\

{\bf Lemma 3.} {\it Let: 1) conditions of lemma 1 take place; 2) $q_N^-$ may not  be stripped-off. Then any 
two formal associated functions of $h^+$ of the same order for the same spectral value $\lambda$, when 
being zero-modes of $q_N^-$, are either simultaneously normalizable at $+\infty$ or simultaneously 
non-normalizable at $+\infty$.
The same takes place at $-\infty$.}\\

We refer the reader for the proofs of the Lemmas 1 -- 3 to the forthcoming Part II.

\section{Interrelation between Jordan cells in SUSY partners}
\hspace*{2ex} The first result on interrelation between Jordan structures of intertwined Hamiltonians and 
on the behavior  of transformation functions at $\pm\infty$ follows from the

{\bf Lemma 4.} {\it Let us assume that: 1) conditions of lemma 1 take place; 2) $\{\varphi^-_{i,j}\}$ and 
$\{\varphi^+_{i,j}\}$ are canonical bases of ${\rm{ker}}\,q_N^-$ and ${\rm{ker}}\,q_N^+$ respectively; 3) 
$q_N^-$ cannot be stripped-off; 4) $k_i$ is an algebraic multiplicity of the eigenvalue $\lambda_i$ of $\bf 
S^+$-matrix (see Ths. 1 and 2). Then for any $i$ and $j$ function $\varphi^-_{i,j}(x)$ is normalizable 
(non-normalizable) at $+\infty$ if and only if $\varphi^+_{i, k_i-j-1}(x)$ is non-normalizable 
(normalizable) at $+\infty$. The same takes place at $-\infty$.}

{\bf Corollary 4.} In order that for the level $\lambda_i$ the Hamiltonian $h^+$ does not have 
eigenfunctions and associated functions normalizable on the whole axis and the Hamiltonian $h^-$ has a 
Jordan cell of multiplicity $\nu_- (\lambda_i)$ spanned by  eigenfunction and associated functions 
normalizable on whole axis (the same number $\nu_- (\lambda_i)$ measures the dimension of the  subspace of 
non-normalizable zero-modes of $q^-$) it is necessary and sufficient that
$$k_i=\max\{k^+_{i\uparrow},k^+_{i\downarrow}\}+\nu_-(\lambda_i) = |k^-_{i\uparrow}-k^-_{i\downarrow}|
+\nu_-(\lambda_i),$$ where $k^\pm_{i\uparrow}$ ($k^\pm_{i\downarrow}$) are numbers of functions 
$\varphi^\mp_{i,j}(x)$ normalizable at $+\infty$ ($-\infty$) (see the previous Sec.). Thus if  there are 
no
 eigenfunctions and associated functions corresponding to the level lambda  $\lambda_i$ of the
initial Hamiltonian $h^+$, normalizable on the whole axis and one wants to get the final Hamiltonian $h^-$ 
with a Jordan cell of rank $\nu_- (\lambda_i)$ spanned by an eigenfunction and associated functions 
normalizable on the whole axis, one must choose transformation functions such that they contain $\nu_- 
(\lambda_i)$ (and no more) associated functions of $h^+$ non-normalizable at both infinities .

A more precise result on interrelation between Jordan structures of intertwined Hamiltonians and on the 
behavior of transformation functions is given in the

\noindent
 {\bf Index Theorem 3:} \ \underline{on relation between Jordan structures
of intertwined Hamiltonians }\\ {\it Let us assume that: 1) conditions of lemma 4 take place; 2) 
$\nu_\pm(\lambda)$ is the algebraic multiplicity of an eigenvalue $\lambda$ of $h^\pm$, {\it  i.e.} the 
number of independent eigenfunctions and associated functions of $h^\pm$ normalizable on the whole axis ; 
3) if $\lambda= \lambda_i$, where $\lambda_i$ is an eigenvalue of $\bf S^\pm$  (see Th. 1), then 
$n_\pm(\lambda_i)$ is a number of normalizable
 functions at both infinities among $\varphi^\mp_{ij}(x)$, $j=0$, \dots,
$k_i-1$ and $n_0(\lambda_i)$ is a number of  functions normalizable only at one of infinities, among 
$\varphi^\mp_{ij}(x)$, $j=0$, \dots, $k_i-1$. Then  the equality
$$\nu_+(\lambda_i)-n_+(\lambda_i)=\nu_-(\lambda_i)-n_-(\lambda_i)$$
takes place for any $i$. Moreover if $n_0(\lambda)>0$ for some $\lambda=\lambda_j$, then for this 
$\lambda_j$
$$\nu_+(\lambda_j)-n_+(\lambda_j)=\nu_-(\lambda_j)-n_-(\lambda_j)=0;$$
and  if $\lambda$ is not an eigenvalue of  $\bf S^\pm$ but  $\lambda\le0$ or Im$\lambda\ne0$, then 
$\nu_+(\lambda)=\nu_-(\lambda) $. }

\section{An example of non-diagonalizable Hamiltonians made by SUSY trans\-for\-ma\-tions}
\def\sh{{\rm{sh}}\,}\def\ch{{\rm{ch}}\,}\def\th{{\rm{th}}\,}
\subsection{SUSY system with Jordan cell of rank 2}
Let us start from the Darboux transformation of the free particle Hamiltonian (which is trivially 
PT-symmetric) , \be h^+=-\partial^2 \ee and build an isospectral Hamiltonian (its SUSY partner) which is 
reflectionless \cite{lahiri,ansok,maydan} due to spectral equivalence to a free particle system, \be 
h^-=-\partial^2-16\alpha^2{{\alpha(x-z)\sh( 
2\alpha(x-x_0))-2\ch^2(\alpha(x-x_0))}\over{[\sh(2\alpha(x-x_0))+ 2\alpha(x-z)]^2}}, \label{hminus}\ee
$$\alpha>0,\qquad x_0\in\mathbb R,\qquad z\in\mathbb C,
\qquad {\rm{Im}}\,z\ne0$$ with the help of the intertwining operators $q_2^\pm$:
$$q_2^\pm h^\mp=h^\pm q_2^\pm,\quad
q_2^-=\partial^2-{{W'(x)}\over{W(x)}}\partial-\alpha^2+{1\over2}{{W''(x)}\over{W(x)}},\quad 
q_2^+=(q_2^-)^t,$$ \be W(x)=\sh(2\alpha(x-x_0))+2\alpha(x-z).\ee If $x_0 ={\rm{Re}}\,z = 0$ the Hamiltonian 
$h^-$ reveals PT-symmetry. Otherwise PT-symmetry is not realized although the energy spectrum remains 
real.

The operator $q_2^-$ can be factorized into two intertwining operators of first order in derivatives, \ba 
q^-_2=q_b^-q_a^-,\quad q_a^-=\partial-\alpha \,\th(\alpha(x-x_0)),\quad 
q_b^-=\partial-{{W'(x)}\over{W(x)}}+\alpha \,\th(\alpha(x-x_0)), \la{fac} \ea with the intermediate 
non-singular Hamiltonian of the ladder construction of Sec. 4, \be 
h_1=-\partial^2-\frac{2\alpha^2}{\ch^2(\alpha(x-x_0))}.\ee

The canonical basis of $q_2^-$ consists of two {\it non-normalizable} functions: 
\be\varphi^-_{0}(x)=\ch(\alpha(x-x_0)),\quad \varphi^-_{1}(x)= 
-{{(x-z)}\over{2\alpha}}\sh(\alpha(x-x_0))+{1\over{4\alpha^2}} \ch(\alpha(x-x_0)), \label{phizero} \ee \be 
h^+\varphi^-_{0}=\lambda_0\varphi^-_{0},\qquad(h^+-\lambda_0)
\varphi^-_{1}=\varphi^-_{0},\quad\lambda_0=-\alpha^2, \quad {\bf S^+} =\left(  \begin{array}{cc} \lambda_0 
& 0\\
1& \lambda_0\end{array} \right) .\ee On the other hand the canonical basis of $q_2^+$ consists of two {\it 
normalizable}  functions: \be \varphi^+_{0}(x)=(2\alpha)^{3/2}{{\varphi^-_{0}(x)}\over{W(x)}}, 
\qquad\varphi^+_{1}(x)=-(2\alpha)^{3/2}{{\varphi^-_{1}(x)}\over{W(x)}}, \label{psi}\ee which form the 
Jordan cell for $h^-$ corresponding to the level $\lambda_0$: \be 
h^-\varphi^+_{0}=\lambda_0\varphi^+_{0},\quad(h^--\lambda_0)
\varphi^+_{1}=\varphi^+_{0},\quad {\bf S^-} =\left(  \begin{array}{cc} \lambda_0 & 0\\
1& \lambda_0\end{array} \right) .\ee

In relation to factorization \gl{fac} one can show that the zero-mode of  $q_b^-$ becomes: \ba 
q_a^-\varphi^-_{1}=-\frac{\sh(\alpha(x-x_0))}{2\alpha} -\frac{(x-z)}{2\ch(\alpha(x-x_0))} \equiv 
-\frac{W(x)}{4\alpha\, \ch(\alpha(x-x_0))} \equiv -\sqrt{\frac{\alpha}{2}}\frac{1}{\varphi^+_{0}}.\ea

In turn the eigenfunctions of $h^-$ for continuous spectrum read: \ba &&\psi(x;k)= - 
\frac{1}{\sqrt{2\pi}(\alpha^2+k^2)}\ q^-_2 e^{ikx} = {1\over\sqrt{2\pi}}\left[1+{{ik}\over{\alpha^2+k^2}} 
{{W'(x)}\over{W(x)}}-{1\over{2(\alpha^2+k^2)}}{{W''(x)}\over{W(x)}} \right]e^{ikx}, \no && k\in\mathbb R, 
\qquad h^-\psi(x;k)=k^2\psi(x;k).\la{psixk1} \ea One can check that eigenfunctions and associated functions 
of $h^-$ obey the relations:
$$\int\limits_{-\infty}^{+\infty}\left( \varphi^+_{0,1}(x)\right)^2\,dx=0,\qquad
\int\limits_{-\infty}^{+\infty}\varphi^+_{0}(x)\varphi^+_{1}(x)\,dx=1,\qquad 
\int\limits_{-\infty}^{+\infty}\varphi^+_{0,1}(x)\psi(x;k)\,dx=0,$$ \be 
\int\limits_{-\infty}^{+\infty}\psi(x;k)\psi(x;-k')\,dx= \delta(k-k') ,\la{(1)}\ee where the last relation 
is understood, as usual, in the sense of distributions.

One can also find that the functions  $\varphi^+_{0}(x)$, $\varphi^+_{1}(x)$ can be obtained by analytical 
continuation of $\psi(x;k)$ in $k$,
$$\lim\limits_{k\to\pm i\alpha}[(k^2+\alpha^2)\psi(x;k)]=\mp\sqrt{\alpha\over\pi}
e^{\mp\alpha x_0}\varphi^+_{0}(x),$$
$$\lim\limits_{k\to\pm i\alpha}\Big[{1\over{2k}}{\partial
\over{\partial k}}\big((k^2+\alpha^2)\psi(x;k)\big)\Big]=\mp\sqrt{\alpha\over\pi}e^{\mp\alpha x_0}\Big[ 
\varphi^+_{1}(x)-{{1\mp2\alpha z}\over{4\alpha^2}}\varphi^+_{0}(x)\Big].$$

For this model resolution of identity made of eigenfunctions and associated functions of $h^-$ can be 
obtained by the conventional Green function method: \be 
\delta(x-x')=\int\limits_{-\infty}^{+\infty}\psi(x;k)\psi(x';-k)\,dk+ 
\varphi^+_{0}(x)\varphi^+_{1}(x')+\varphi^+_{1}(x)\varphi^+_{0}(x'),\la{razl1}\ee or in the operator form, 
\be I=\int\limits_{-\infty}^{+\infty}|\psi,k\rangle\langle\tilde\psi,k|\,dk+ 
|\psi_0\rangle\langle\tilde\psi_0|+|\psi_1\rangle\langle\tilde\psi_1|,\ee where the Dirac notations have 
been used: \be\langle x|\psi,k\rangle=\psi(x;k),\qquad\langle x|\tilde\psi,k\rangle=\psi^*(x;-k),\ee 
\be\langle x|\psi_{0,1}\rangle=\varphi^+_{0,1}(x),\qquad \langle x|\tilde\psi_{0,1}\rangle=\left( 
\varphi^+_{1,0}(x) \right)^*,\ee \be h^{-\dagger}|\tilde\psi_1\rangle=\lambda_0|\tilde\psi_1\rangle,\qquad 
(h^{-\dagger}-\lambda_0)|\tilde\psi_0\rangle=|\tilde\psi_1\rangle,\qquad
h^{-\dagger}|\tilde\psi,k\rangle=k^2|\tilde\psi,k\rangle \ee We stress that $|\tilde\psi_{0,1}\rangle$ are 
analogs of $|\tilde\psi_{n,a,i}\rangle$ from Sec.~2. In this notations the biorthogonal relations take the 
form, \be \langle\tilde\psi_{j}|\psi_{k} \rangle=\delta_{jk},\quad \langle\tilde\psi,k|\psi_{0,1}\rangle= 
\langle\tilde\psi_{1,0}|\psi,k\rangle=0,\quad \langle\tilde\psi,k|\psi,k'\rangle= \delta(k-k').\ee 
Accordingly, the spectral decomposition of $h^-$ can be easily derived, \be 
h^-=\int\limits_{-\infty}^{+\infty}k^2|\psi,k\rangle\langle\tilde\psi,k| 
\,dk-\alpha^2|\psi_0\rangle\langle\tilde\psi_0|- \alpha^2|\psi_1\rangle\langle\tilde\psi_1|+ 
|\psi_0\rangle\langle\tilde\psi_1|.\ee

Let's remind that, for a given $\lambda = \lambda_i$, $k_i$  is the number of zero-modes of $q^\mp_2$, 
$k^\pm_{i\uparrow,\downarrow}$ are the numbers of zero-modes normalizable at one of $\pm\infty$ labeled by 
$\uparrow, \downarrow$, $\nu_\pm(\lambda)$ are the numbers of eigenfunctions and associated functions of 
$h^\pm$ normalizable on the whole axis, $n_\pm(\lambda)$ are
 the numbers of  eigenfunctions and associated functions among zero-modes $\varphi^\mp_{i,j}$ normalizable 
on the whole axis
and $n_0(\lambda)$ is the number of eigenfunctions and associated functions among zero-modes 
$\varphi^\mp_{i,j}$ normalizable only at one end. They are defined in Sec.~5,~6 . For this model they take 
the following particular values ($i = 0$):
$$k_0=2,\qquad k^+_{0\uparrow,\downarrow}=0,\qquad\nu_+(\lambda)\equiv0,\qquad 
k^-_{0\uparrow,\downarrow}=2,\qquad
\nu_-(\lambda)=\begin{cases}2,& \lambda=\lambda_0,\\0,&\lambda\ne\lambda_0,\end{cases}$$
$$n_+(\lambda)\equiv0; \qquad
n_0(\lambda)\equiv0,\qquad n_-(\lambda)=\begin{cases}2,& 
\lambda=\lambda_0,\\0,&\lambda\ne\lambda_0,\end{cases},$$ and the Index Theorem holds,
$$\nu_+(\lambda_j)-n_+(\lambda_j)=\nu_-(\lambda_j)-n_-(\lambda_j)=0 . $$
\subsection{Coalescence of two levels}
The Hamiltonian $h^-$ with Jordan cell for bound state \gl{hminus} is a particular limiting case of the 
Hamiltonian $h^-$ with two  non-degenerate bound states (of algebraic multiplicity~1). The former 
corresponds to the confluent case of the latter one. One again starts from the free particle Hamiltonian,
$$h^+=-\partial^2$$
and obtains its SUSY partner
$$h^-=-\partial^2-16\alpha^2\times$$ $${{
{{\alpha^2+\beta^2}\over{2\alpha\beta}} 
\sh(2\alpha(x-x_0))\sh(2\beta(x-z))\!\!-\!\!2\ch^2(\alpha(x-x_0))\ch(2\beta(x-z))\!\!+\!\! 
2\sh^2(\beta(x-z))} \over{[\sh(2\alpha(x-x_0))+{\alpha\over\beta}\sh(2\beta(x-z))]^2}}$$ by intertwining 
with the operators
$$q_2^-=\partial^2-{{W'(x)}\over{W(x)}}\partial-(\alpha^2+\beta^2)+{1\over2}{{W''(x)}\over{W(x)}} = 
(q_2^+)^t ,$$
where
$$W(x)=\sh(2\alpha(x-x_0))+{\alpha\over\beta}\sh(2\beta(x-z)),$$
$$x_0\in\mathbb R,\qquad  {\rm{ Im}}\, z\ne0, \qquad  \alpha>0\quad ({\rm {or}}\,\,\, -i\alpha>0),\qquad
0\le\beta<{\pi\over{2{\rm{Im}}\, z}}.$$ In the case $\beta\ne0$, $\beta\ne\alpha$ the canonical basis of $ 
q_2^-$ zero-modes (transformation functions) consists of
$$\varphi_{\pm\beta}(x)=\ch(k_{\pm\beta}(x-\xi_{\pm\beta})),\qquad k_{\pm\beta}=\alpha\pm\beta,
\quad\xi_{\pm\beta}={{\alpha x_0\pm\beta z}\over{\alpha\pm\beta}},$$ so that
$$h^+\varphi_{\pm\beta}=\lambda_{\pm\beta}\varphi_{\pm\beta},\qquad\lambda_{\pm\beta}=-k_{\pm\beta}^2=
-(\alpha\pm\beta)^2.$$ One can check that  in the case $\beta\ne0$ the function $W(x)$ is a Wronskian of 
$\varphi_{+\beta}(x)$ and $\varphi_{-\beta}(x)$ divided by $\beta$ and in the case~$\beta=0$ it is a 
product of $-4\alpha$ and of the Wronskian of $\varphi^-_{0}(x)$ and $\varphi^-_{1}(x)$.

For $\beta=0$ both functions $\varphi_{\pm\beta}(x)$ coincide with $\varphi^-_{0}(x)$ from \gl{phizero}, 
and as well $\lambda_{\pm\beta}=\lambda_0$.
 In this case $\varphi^-_{1}(x)$ is a linear combination of
$\varphi_{+\beta}(x)$ and $\varphi_{-\beta}(x)$ in the following sense,
$$\varphi^-_{1}={{{\partial\over{\partial\beta}}(\varphi_{+\beta}-\varphi_{-\beta})}\over
{{\partial\over{\partial\beta}}(\lambda_{+\beta}-\lambda_{-\beta})}}\bigg\vert_{\beta=0}+ 
{1\over{4\alpha^2}}\varphi^-_0.$$

In the case $\beta\ne0$ the canonical basis of zero-modes of $q^+_2$ consists of
$$\tilde\psi_{\pm\beta}(x)={{\varphi_{\mp\beta}(x)}\over {W(x)}}.$$
One can prove that \be 
\int\limits_{-\infty}^{+\infty}\tilde\psi^2_{\pm\beta}(x)\,dx=\mp{\beta\over{2\alpha(\alpha\pm\beta)}} 
\la{(3)}\ee (this formula is valid also in the case $\beta=0$). The fact that the integrals \gl{(3)} vanish 
for $\beta=0$ is in line with \gl{(1)}. It follows from \gl{(3)} that the normalized eigenfunctions of 
$h^-$ have the form
$$\psi_{+\beta}(x)=\sqrt{2}i\alpha\sqrt{{1\over\beta}+{1\over\alpha}}{{\varphi_{-\beta}(x)}\over{W(x)}},\quad
\psi_{-\beta}(x)=\sqrt{2}\alpha\sqrt{{1\over\beta}-{1\over\alpha}}{{\varphi_{+\beta}(x)}\over{W(x)}},\quad 
h^-\psi_{\pm\beta}= \lambda_{\pm\beta}\psi_{\pm\beta}.$$

The eigenfunctions of $h^-$ for continuous spectrum read, 
\ba&&\psi(x;k)={{[\alpha^2+\beta^2+k^2+ik{{W'(x)}\over{W(x)}}-{1\over2}{{W''(x)} 
\over{W(x)}}]e^{ikx}}\over{\sqrt{2\pi}\sqrt{(k^2+\alpha^2+\beta^2)^2-4\alpha^2\beta^2}}},\nonumber\\&&k\in\mathbb 
R,\qquad h^-\psi(x;k)=k^2\psi(x;k),\la{psixk2}\ea where the branch of 
$\sqrt{(k^2+\alpha^2+\beta^2)^2-4\alpha^2\beta^2}$ can be defined by the condition
$$\sqrt{(k^2+\alpha^2+\beta^2)^2-4\alpha^2\beta^2}=k^2+o(k^2),\qquad k\to\infty$$
in the plane with cuts, linking branch points, situated in the upper (lower) half-plane.

One can prove the following limits,
$$\lim\limits_{k\to\pm i(\alpha+\beta)}[\sqrt{(k^2+\alpha^2+\beta^2)^2-4\alpha^2\beta^2}\psi(x;k)]=
\pm {{2i\alpha\beta}\over\sqrt\pi}\sqrt{{1\over\beta} +{1\over\alpha}}\,e^{\mp(\alpha x_0+\beta 
z)}\psi_{+\beta}(x),$$
$$\lim\limits_{k\to\pm i(\alpha-\beta)}[\sqrt{(k^2+\alpha^2+\beta^2)^2-4\alpha^2\beta^2}\psi(x;k)]=
\mp {{2\alpha\beta}\over\sqrt\pi}\sqrt{{1\over\beta} -{1\over\alpha}}\,e^{\mp(\alpha x_0-\beta 
z)}\psi_{-\beta}(x) .$$ In the limiting case $\beta=0$ the eigenfunction $\varphi_0^+(x)$ and the 
associated function $\varphi_1^+(x)$ of $h^-$ (see Eqs.\gl{psi}) can be derived from $\psi_{\pm\beta}(x)$ 
,
$$\varphi_0^+(x)=-2i\sqrt\alpha\lim\limits_{\beta\to0}[\sqrt\beta\psi_{+\beta}(x)]=2\sqrt\alpha\lim\limits_{\beta\to0}[\sqrt\beta\psi_{-\beta}(x)],$$
$$\varphi_1^+(x)=2\sqrt{\alpha}\lim\limits_{\beta\to0}{{{\partial\over{\partial\beta}}\big[\sqrt\beta\big(\psi_{-\beta}(x)+i\psi_{+\beta}(x)\big)\big]}
\over{{\partial\over{\partial\beta}}(\lambda_{+\beta}-\lambda_{-\beta})}}.$$

Resolution of identity in the case $\beta\ne0$ takes the form: 
\be\delta(x-x')=\psi_{+\beta}(x)\psi_{+\beta}(x')+\psi_{-\beta}(x)\psi_{-\beta}(x')+ 
\int\limits_{-\infty}^{+\infty}\psi(x;k)\psi(x;-k)\,dk\la{decom2}\ee and one can show that in the case 
$\alpha>0$
$$\lim\limits_{\beta\to 0}[\psi_{+\beta}(x)\psi_{+\beta}(x')+\psi_{-\beta}(x)\psi_{-\beta}(x')]=
\varphi_0^+(x)\varphi_1^+(x')+\varphi_1^+(x)\varphi_0^+(x')$$ (cf. with \gl{razl1}).

One can also check that the biorthogonal relations take place: \be 
\int\limits_{-\infty}^{+\infty}\psi_{+\beta}(x)\psi_{-\beta}(x)\,dx=0,\quad 
\int\limits_{-\infty}^{+\infty}\psi_{\pm\beta}(x)\psi(x;k)\,dx=0,\quad 
\int\limits_{-\infty}^{+\infty}\psi(x;k)\psi(x;-k')\,dx= \delta(k-k').\la{biort2}\ee
\subsection{Symmetry operators}
For the Hamiltonian $h^-$ of Subsec. 1 and 2 there exists the antisymmetric symmetry operator of the fifth 
order
$$R_5=q_2^-\partial q_2^+,\qquad R_5h^-=h^-R_5,\qquad R_5^t=-R_5 .$$
In the case of PT-symmetry,  $x_0 ={\rm{Re}}\,z = 0$ it anticommutes with PT-reflection $\theta_PT$, $R_5 
\theta_PT = - \theta_PT R_5$.

The  wave function $\varphi^+_{0}(x)$, the associated function $\varphi^+_{1}(x)$ and the wave function of 
zero-energy bound state 
$$\qquad\qquad\psi(x;0)={1\over\sqrt{2\pi}}\Big[1-{1\over{2\alpha^2}}{{W''(x)}\over{W(x)}}\Big]$$
are zero-modes of $R_5$ because of $\varphi^+_{0,1}(x)$ are zero-modes of $q_2^+$ and 
$q_2^+\psi(x;0)={\rm{Const}}$. In Subsec. 2 (in the case $\beta\ne0$) the wave functions 
$\psi_{+\beta}(x)$, $\psi_{-\beta}(x)$ and the wave function of zero-energy bound state
$$\psi(x;0)={{[\alpha^2+\beta^2-{1\over2}{{W''(x)}\over{W(x)}}]}\over{\sqrt{2\pi}(\alpha^2-
\beta^2)}}$$ are zero-modes of $R_5$ because of $\psi_{\pm\beta}(x)$ are zero-modes of $q_2^+$ and 
$q_2^+\psi(x;0)={\rm{Const}}$. In both cases the  spectrum of $\bf S$-matrix of the symmetry operator (an 
analogue of $\bf S^+$ from Th.~1) consists of $0$ and $-(\alpha\pm\beta)^2$
.

We notice that if $k\ne0$ then $\psi(x;k)$ and $\psi(x;-k)$  are linearly independent
wave functions for the energy level $E=k^2$. Simultaneously, $\psi(x;k)$ is an eigenfunction of the 
symmetry operator:
$$R_5\psi(x;k)=ik(k^2+\alpha^2)^2\psi(x;k),$$
or
$$R_5\psi(x;k)=ik[(k^2+\alpha^2+\beta^2)^2-4\alpha^2\beta^2]\psi(x;k),$$
Thus the zeroes of the eigenvalue of $R_5$  are related  to eigenvalues of $\bf S$-matrix of $R_5$ in 
accordance to \cite{ansok}.

\section{Conclusions and perspectives: peculiarities of non-Hermitian Hamiltonians with
continuous spectrum} In our paper we have investigated a bound state part of the spectrum for the class of 
potentials among which one can find also those ones with continuum spectrum.  The choice of this class has 
allowed to keep all the analysis well below a possible continuum threshold. For this part of the spectrum  
the relationship  between Jordan cells of SUSY partner Hamiltonians is firmly controlled by the Index 
Theorem 3 and it was well illuminated by an exactly solvable system with two coalescing bound states. We 
remind that the rigorous proofs of all new Lemmas and the Theorem 3 are postponed to the second part of 
this paper \cite{sokolov}.

Meantime  the approaching to the continuum threshold yields more subtle  problems with normalizable eigen- 
and associated functions in continuum which may have zero binorm. As a consequence it may cause serious 
problems with the resolution of identity . This interesting problem we investigated in \cite{sokancan}. 
Here  we would like only to draw attention to a class of models where the continuous spectrum is involved  
and elaborate the resolution of identity.

Let us  consider the model Hamiltonians \be h^+=-\partial^2,\qquad 
h^-=-\partial^2+{2\over{(x-z)^2}},\quad{\rm{Im}}\, z\ne0, \ee intertwined by the first-order operators 
$q_1^\pm$: \be h^\pm q_1^\pm=q_1^\pm h^\mp,\qquad q_1^\pm=\mp\partial-{1\over{x-z}}.\ee If ${\rm{Re}}\,z = 
0$ these Hamiltonians are PT-symmetric and respectively their eigenfunctions possess definite PT- 
parities.

The eigenfunctions of $h^-$ of continuous spectrum can be found in the form, \be 
\psi(x;k)={1\over\sqrt{2\pi}}\left[1-{1\over{ik(x-z)}}\right]e^{ikx},\qquad k\in\mathbb R\backslash\{0\}, 
\qquad h^-\psi(x;k)=k^2\psi(x;k). \la{cont3}\ee In addition, there is a normalizable eigenfunction of $h^-$ 
on the lower end of continuous spectrum:\be 
\psi_0(x)={1\over{(x-z)}}=-\sqrt{2\pi}\lim\limits_{k\to0}[ik\psi(x;k)],\qquad h^-\psi_0=0.\ee Isospectral 
relations between $h^+$ and $h^-$ take the form, \be 
q^-\Big[{e^{ikx}\over\sqrt{2\pi}}\Big]=ik\psi(x;k),\quad k\ne0,\qquad q^- \Big[{1\over\sqrt{2\pi}}\Big]= 
{-1\over\sqrt{2\pi}}\psi_0(x),\quad k=0;\ee \be q^+\psi(x;k)=-ik\Big[{e^{ikx}\over\sqrt{2\pi}}\Big],\quad 
k\ne0,\qquad q^+\psi_0=0,\quad k=0.\ee The eigenfunctions of $h^-$ satisfy the relations of 
biorthogonality,
\begin{equation}\int\limits_{-\infty}^{+\infty}\psi_0^2(x)\,dx=0,\qquad
\int\limits_{-\infty}^{+\infty}\psi_0(x)\psi(x;k)\,dx=0.\la{1}\end{equation} Resolution of identity made of 
eigenfunctions of $h^-$ can be built as follows,
\begin{equation}\delta(x-x')=\int\limits_{\cal L}\psi(x;k)\psi(x';-k)\,dk,
\la{2}\end{equation} where the contour $\cal L$ must be a proper integration path in the complex $k$ plane 
which allows to regularize the singularity in \gl{cont3} for $k = 0$ circumventing it from up or from down. 
To reach an adequate definition of resolution of identity one can instead use the  Newton--Leibnitz formula  
and rewrite \gl{2} in the form
$$ \delta(x-x')=\Big(\int\limits_{-\infty}^{-\varepsilon}+
\int\limits_{\varepsilon}^{+\infty}\Big)\psi(x;k)\psi(x';-k)\,dk$$
\begin{equation}- {{\psi_0(x)\psi_0(x')}\over{\pi\varepsilon}}+{{\sin\varepsilon
(x-x')}\over{\pi(x-x')}}+{{2\sin^2({\varepsilon\over2}(x-x'))} 
\over{\pi\varepsilon(x-z)(x'-z)}},\qquad\varepsilon>0.\la{3}\end{equation} One can show \cite{sokancan} 
that the limit of the 3rd term of the right side of \gl{3} (as a distribution) under 
$\varepsilon\downarrow0$ is zero for any test function from $C_{\mathbb R}^\infty\cap L^2({\mathbb R})$ but 
the limit of the last term of the right side of \gl{3} under $\varepsilon\downarrow0$ is zero only for test 
functions from $C_{\mathbb R}^\infty\cap L^2({\mathbb R};|x|^\gamma)$, $\gamma>1$. Thus for test functions 
from $C_{\mathbb R}^\infty\cap L^2({\mathbb R};|x|^\gamma)$, $\gamma>1$ resolution of identity can be 
reduced to,
\begin{equation}\delta(x-x')=\lim\limits_{\varepsilon\downarrow0}\bigg[
\Big(\int\limits_{-\infty}^{-\varepsilon}+ 
\int\limits_{\varepsilon}^{+\infty}\Big)\psi(x;k)\psi(x';-k)\,dk- 
{{\psi_0(x)\psi_0(x')}\over{\pi\varepsilon}}\bigg]\la{4}\end{equation} and for test functions from 
$C_{\mathbb R}^\infty\cap L^2({\mathbb R})$ to, \ba 
\delta(x-x')&=&\lim\limits_{\varepsilon\downarrow0}\bigg\{ \Big(\int\limits_{-\infty}^{-\varepsilon}+ 
\int\limits_{\varepsilon}^{+\infty}\Big)\psi(x;k)\psi(x';-k)\,dk\no &&- {1\over{\pi\varepsilon}}\Big[1- 
2\sin^2\big({\varepsilon\over2}(x-x')\big)\Big]{\psi_0(x)\psi_0(x')}\bigg\}.\la{5}\ea Decomposition \gl{4} 
seems to have a more natural form than \gl{5}, but its right side obviously cannot reproduce the {\it 
normalizable} eigenfunction
$$\psi_0(x)\not\in C_{\mathbb R}^\infty\cap L^2({\mathbb R};|x|^\gamma), \qquad
\gamma>1$$ because of the orthogonality relations \gl{1}. One can show that \be 
\lim\limits_{\varepsilon\downarrow0} 
\int\limits_{-\infty}^{+\infty}{2\over{\pi\varepsilon}}{\sin^2\big({\varepsilon\over2}(x-x')\big)} 
\psi_0^2(x)\psi_0(x')\,dx=\lim\limits_{\varepsilon\downarrow0}[e^{-i\varepsilon 
x'}\psi_0(x')]=\psi_0(x').\ee Thus it is the 3rd term in the right side of \gl{5} that provides the 
opportunity to reproduce $\psi_0(x)$ and thereby to complete the resolution of identity.

The spectral decomposition of $h^-$ in this case reads \be 
h^-=\int\limits_{-\infty}^{+\infty}k^2|\psi,k\rangle\langle\tilde\psi, k|\,dk,\ee where
$$\langle x|\psi,k\rangle=\psi(x;k),\qquad \langle x|\tilde\psi,k\rangle=\psi^*(x;-k),
\qquad h^{-\dagger}|\tilde\psi,k\rangle=k^2|\tilde\psi,k\rangle.$$

 For the Hamiltonian $h^-$ of this model there is an antisymmetric symmetry operator of the
3rd order
$$R_3=q_1^-\partial q_1^+=-\partial^3+{3\over{(x-z)^2}}\partial-{3\over{(x-z)^3}},$$ $$
 R_3h^-=h^-R_3,\quad R_3^t=-R_3,\quad R_3\psi(x;k)=ik^3\psi(x;k) .$$

We notice that this model is a limiting case of the example of Subsec. 7.1 for $\alpha\to0$ where $z$ must 
be taken as a half sum of $x_0$ and $z$.

One can generalize this model constructing  the  Hamiltonian (by intertwining with the Hamiltonian of a 
free particle)
$$h^-=-\partial^2 +{{n(n+1)}\over{(x-z)^2}},$$
for which there is  a Jordan cell, spanned by $\big[{{n+1}\over2}\big]$ normalizable eigenfunction and 
associated functions:
$$h\psi_0=0,\qquad h\psi_j=\psi_{j-1},\qquad j=0,\ldots, \big[{{n-1}\over2}\big],
\qquad\psi_j(x)={{(2(n-j)-1)!!}\over{(2j)!!(2n-1)!!(x-z)^{n-2j}}} ,$$ at the bottom of continuous spectrum 
. All these functions are mutually biorthogonal having also zero binorm. The problem of a correct 
resolution of identity seems to be solvable in the same way as for $n=1$ . However the rigorous analysis is 
postponed to a forthcoming paper.
\section{Acknowledgements}
The work of A.A. and A.S. was supported by Grant RFBR 06-01-00186-a and  by  Programs RNP 2.1.1.1112 and 
LSS-5538.2006.2.  A.Sokolov was partially supported by the INFN grant.

\end{document}